# Temporal fluctuations in excimer-like interactions between π-conjugated chromophores

*Thomas Stangl[1], Philipp Wilhelm[1], Daniela Schmitz[2], Klaas Remmerssen[2], Sebastian Henzel[2], Sigurd Höger[2], Jan Vogelsang[1],\*, John M. Lupton[1]*

[1]Institut für Experimentelle und Angewandte Physik, Universität Regensburg, Universitätsstrasse 31, 93053 Regensburg, Germany

[2]Kekulé-Institut für Organische Chemie und Biochemie der Universität Bonn, Gerhard-Domagk-Str. 1, 53121 Bonn, Germany

AUTHOR INFORMATION

**Corresponding Author**

\* Jan Vogelsang, Institut für Experimentelle und Angewandte Physik, Universität Regensburg, Universitätsstrasse 31, 93053 Regensburg, Germany, E-Mail: <jan.vogelsang@physik.uni-regensburg.de>

ABSTRACT

Inter- or intramolecular coupling processes between chromophores such as excimer formation or H- and J-aggregation are crucial to describing the photophysics of closely packed films of conjugated polymers. Such coupling is highly distance dependent, and should be sensitive to both fluctuations in the spacing between chromophores as well as the actual position on the chromophore where the exciton localizes. Single-molecule spectroscopy reveals these intrinsic fluctuations in well-defined bi-chromophoric model systems of cofacial oligomers. Signatures of interchromophoric interactions in the excited state – spectral red-shifting and broadening, and a slowing of photoluminescence decay – correlate with each other but scatter strongly between single molecules, implying an extraordinary distribution in coupling strengths. Furthermore, these excimer-like spectral fingerprints vary with time, revealing intrinsic dynamics in the coupling strength within one single dimer molecule, which constitutes the starting point for describing a molecular solid. Such spectral sensitivity to sub-Angstrom molecular dynamics could prove complementary to conventional FRET-based molecular rulers.

**TOC GRAPHICS**

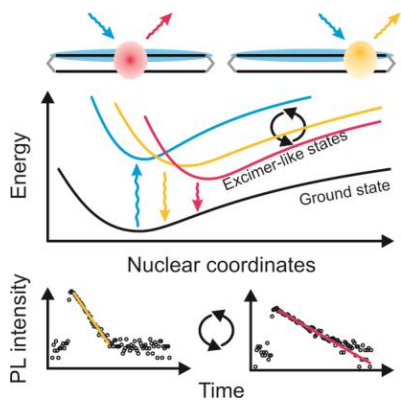

The fact that π-conjugated materials are packed closely together once they find use in optoelectronic devices[1-3] stimulates the demand for a detailed microscopic picture of interchromophoric coupling.[4,5] Based on the understanding of spectral splitting in statistical dimer pairs,[6] a broad spectrum of reports describe theoretically and confirm experimentally the existence of interchromophoric coupling in conjugated polymers[6-12] by utilizing ensemble-based spectroscopic methods. Such coupling can occur in addition to FRET-like inter- and intrachain energy transfer.[13] However, it is not always straightforward to identify spectral signatures of such coupling unambiguously in ensemble spectroscopy because chemical defects[14,15] or strong spectral diffusion due to intrachromophoric fluctuations in transition energy[16] may give rise to similar spectroscopic signatures such as spectral broadening or lifetime changes. Coupling is primarily described in the framework of excimeric states,[7,9,11] H-aggregation,[17] J-aggregation, π-π stacking, a combination thereof[18-20] or more generally as physical dimers.[6] Excimeric states only require coupling in the excited state, whereas H- or J-aggregation implies overlap of the electron wave function of both monomers in the ground and excited states. Since coupling efficiency depends critically on distance, tiny variations in the location of the primary exciton can dramatically impact the spectroscopic observable as was already suggested previously by Schindler et al. on conjugated polymers.[21] Ensemble techniques overlook this variability, providing a time and space average. Additionally, the substantial electronic and morphological heterogeneity of conjugated polymers[1-3,22] results in such a degree of complexity that intrinsic structure-function relationships are often masked by disorder in the bulk. This uncertainty arises because the chromophores themselves can differ in size,[23] morphology,[24] energy levels[25] and orientation with respect to each other.[26]

A promising Ansatz to circumvent these obstacles is a bottom-up approach using well-defined model systems of conjugated polymers in combination with single-molecule spectroscopy.[27-33] This strategy has not been pursued extensively in the context of cofacial excimeric states, although a few examples of bi- and tri-chromophoric single-molecule J-aggregate models exist, based mostly on perylenes.[31,34-37] A single-molecule investigation should lead to new insights regarding the heterogeneity and dynamics of interchromophoric coupling. Here, we investigate the impact of fluctuations in exciton localization and molecular structure in extended chromophores on the formation of excimer-like intramolecular excitations. We find that the formation of a interchromophic emissive excited state species is not a strictly deterministic process.

We approach the problem with covalently bound dimers of π-conjugated oligomers. Fig. 1 shows the chemical structure of the individual oligomer. The complete chemical structures of the model dimers are given in Fig. S1 and ensemble absorption and photoluminescence (PL) emission spectra are shown in Fig. S2.[29] As the active entities we chose chromophores based on poly(phenylene-ethynylene) (PPE) since this material shows a strong red-shift and loss of vibronic structure in PL between solution and film,[38] which is consistent with the formation of an excimer-like state.[7] In addition, we do not observe a blue shift for the absorption spectra of the 4.6 Å dimer compared to the monomer (shown in Fig. S2), concluding that there is no coupling in the ground state, which would be present in a physical dimer (H-aggregate). The spacing between the two π-conjugated oligomers is varied by a molecular clamp structure from 21 Å down to 4.6 Å. The latter value is comparable to the distance between chromophores in conjugated polymer films packed in an ordered structure, for which signatures of excitonic coupling have been reported.[17,39]

We begin by searching for spectroscopic signatures of excitonic dimerization in the four dimers on the single-molecule level. Interchromophoric coupling should manifest itself in a red-shift of the emission spectrum, a decrease of the 0-0 to 0-1 peak ratio up to a total disappearance of vibronic structure in a perfect excimer, and an increase of the PL lifetime ($\tau_{PL}$) due to a decrease of the radiative rate.[17,40] The molecules were dissolved in a poly(methyl-methacrylate) (PMMA) / toluene solution and diluted to single molecule concentrations (~$10^{-13}$ M). The solution was spin-coated on cleaned thin glass coverslips to obtain a 50-100 nm thick PMMA film with uniformly distributed single molecules embedded into it (see ref.[28] for details of sample preparation). The films were scanned by a custom-designed confocal microscope (see *Supporting Information* and ref.[28] for details) to identify the locations of single molecules. This information was used to position one single molecule at a time into the diffraction-limited excitation spot at 3.06 eV photon energy excitation. The PL was passed through a 30/70 beam-splitter onto an avalanche photodiode connected to a time-correlated single-photon counting unit, and onto a spectrometer with a CCD camera, respectively, enabling $\tau_{PL}$ and the PL spectrum to be measured at the same time for a single molecule. Each molecule was recorded for a duration of 4 s and discarded if the overall PL intensity dropped by more than 20 % with respect to the initial value. Typical PL intensity traces with the corresponding PL decays and spectra for two different dimers are shown in Fig. S3. $\tau_{PL}$ values were extracted from single-exponential fits to the PL decay (see Fig. S3 c, g for details). The PL peak position ($E_{0-0}$) was obtained by fitting the spectral shape with two Gaussians.

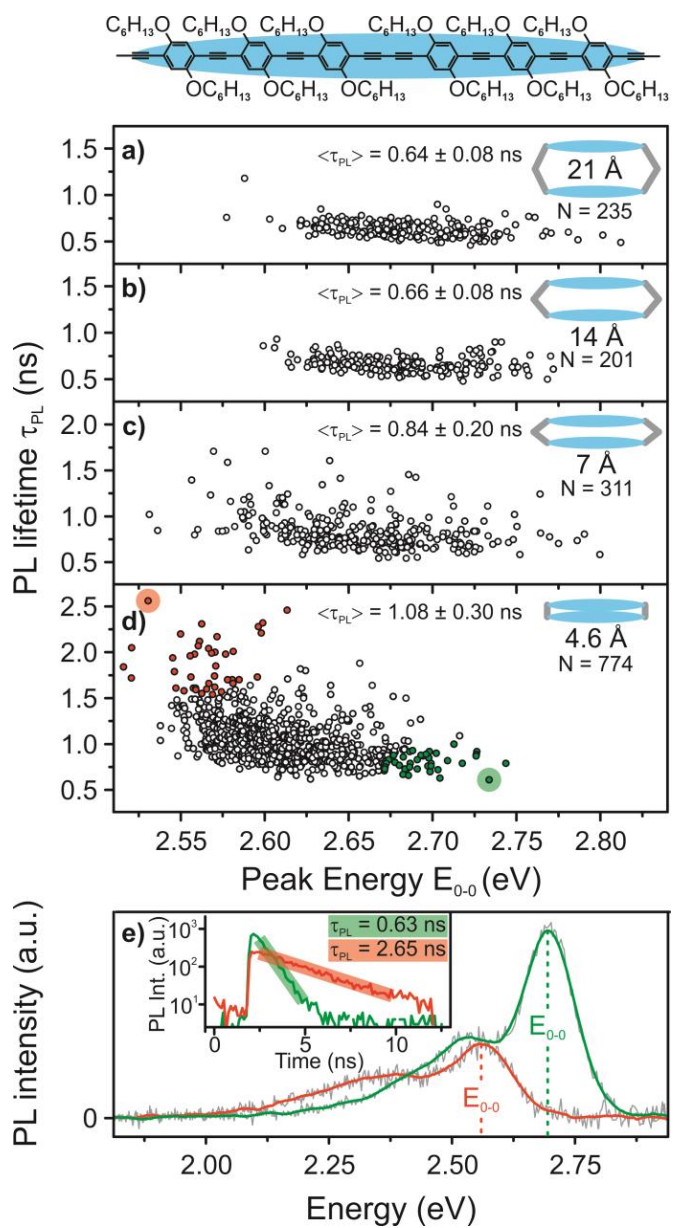

**Figure 1.** Formation of an intramolecular excimer-like state in model systems of π-conjugated oligomer dimers revealed by the correlation between PL lifetime ($\tau_{PL}$) and spectral peak position ($E_{0-0}$). The chemical structure of the model π-conjugated oligomer (5 nm in length) is given above. Panels (a-d) show the relation between $\tau_{PL}$ and $E_{0-0}$ for differently spaced oligomer dimers. A trend to longer $\tau_{PL}$ and lower $E_{0-0}$ values is apparent for the narrowest dimers (d). The average $\tau_{PL}$ and the corresponding standard deviation are stated, as is the number of molecules

measured for each sample. (e) Summed PL spectra from the sub-populations denoted in red and blue in panel (d). The inset shows the PL lifetime decays for the molecule with the highest (marked in panel (d) with a red circle) and lowest (marked green) PL lifetime, respectively.

Fig. 1 a-d) shows the scatter of $\tau_{PL}$ and $E_{0-0}$ peak energies between different single molecules, depending on dimer spacing. $E_{0-0}$ values scatter by ~ 0.2 eV, mainly due to the dielectric heterogeneity in the sample.[41] This scatter of $E_{0-0}$ can be seen in all samples, whereas $\tau_{PL}$ values are narrowly distributed with a standard deviation $\sigma(\tau_{PL})$ of 0.08 ns in the 21 Å and 14 Å dimers (Fig. 1 a, b). A slight tendency for higher $\tau_{PL}$ and lower $E_{0-0}$ values is observed in the 7 Å dimer, where $\sigma(\tau_{PL})$ increases to 0.2 ns (Fig. 1c). In contrast, a clear correlation between $\tau_{PL}$ and $E_{0-0}$ arises for the smallest 4.6 Å dimer (Fig. 1d). Whereas the distribution of $E_{0-0}$ values broadens only slightly up to a maximum spread of 0.24 eV, the $\tau_{PL}$ values scatter widely from 0.6 ns up to 2.6 ns with a standard deviation of 0.3 ns. We note that the chemical structure of the clamps differs slightly for the 4.6 Å dimer in comparison to the larger dimers. For the closest dimers anthracene units had to be used, which are conjugated, whereas non-conjugated *para*-phenylenes could be used for larger spacing. Some electronic delocalization into the anthracene clamp unit may be anticipated for the closest dimers. To test the possible impact of such delocalization on the scatter of $\tau_{PL}$ and $E_{0-0}$ peak energies, an alternative structure was chosen for 4.6 Å spacing, which contains non-conjugated xanthene clamps (shown in Fig. S1 e). Here, π-conjugation is definitely disrupted in the clamps. Consequently, a slight red-shift in absorption and PL emission is observed for the anthracene clamp dimers compared to the xanthene dimers (see Fig. S2 d). However, Fig. S4 shows that the same scatter of $\tau_{PL}$ versus $E_{0-0}$ is found for both dimers of 4.6 Å spacing, irrespective of the clamp used. We conclude that the correlation between $E_{0-0}$ and $\tau_{PL}$ must arise due to local variations in interchromophoric coupling, i.e. in excimer-like states, and is not perturbed by electronic coupling to the clamp. The average spectral shift from the monomer to the most closely spaced dimer is 400 cm$^{-1}$. This value provides a measure of the "intramolecular" (i.e. interchromophoric) electronic coupling strength.

Single-molecule spectroscopy allows us to inspect sub-populations within the 4.6 Å dimer. In Fig. 1 e), we summed up spectra for 40 molecules with the highest $\tau_{PL}$ and lowest $E_{0-0}$ values (marked red in Fig. 1 d) and compared this sum to the 40 molecules with the lowest $\tau_{PL}$ and highest $E_{0-0}$ values (marked green in Fig. 1 d). The inset provides an example of the PL decay transient for the two single molecules with the highest (red curve, $\tau_{PL}$ = 2.65 ns) and lowest PL lifetime (green curve, $\tau_{PL}$ = 0.63 ns), respectively. A clear reduction of the 0-0 to 0-1 peak ratio is seen together with a slight broadening of the vibronic peak for the population with the highest $\tau_{PL}$ values, which is consistent with an increase in interchromophoric coupling, i.e. the formation of a weak excimer-like state.[11] In addition, the average integrated intensity of the red spectrum is three times lower than that of the blue spectrum, since the decrease in radiative rate for the intramolecular excited state results in an overall reduction in fluorescence yield provided a non-radiative decay pathway is present. For further clarification, Fig. S5 d) demonstrates the correlation between the shift of the 0-0 peak with increasing 0-0 to 0-1 peak ratio for four single 4.6 Å dimers. This correlation is not seen in the 21 Å dimers (see Fig. S5 c). We note that the close proximity of the two oligomers could also lead to bending in the individual oligomers themselves, due to steric interactions, which could change the spectroscopic signatures without requiring any interchromophoric coupling.[24,42] To test for bending in the 4.6 Å dimers, we compared the model systems with respect to the orientation of their overall transition dipole moment, which is determined by excitation polarization anisotropy.[26] In this experiment, the excitation beam is linearly polarized in the xy-plane of the sample and the polarization is rotated while the PL response is recorded (for further information see ref.[28]). No difference is found between the dimers with regards to the excitation polarization anisotropy (see Fig. S6). All dimers are universally highly anisotropic, demonstrating that the optically active oligomer units all have the same shape, independent of spacing.

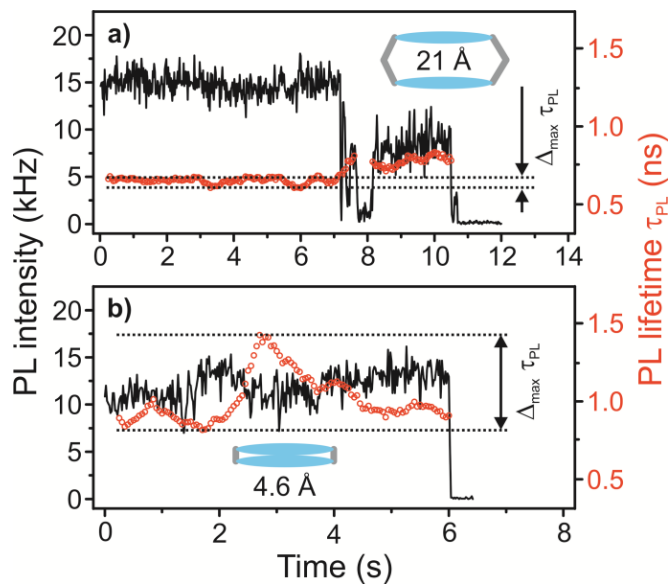

**Figure 2.** Fluctuations in interchromophoric coupling strength in the excited state revealed by tracking the PL lifetime as a function of time for one single molecule. (a) PL intensity transient of a 21 Å dimer obtained by confocal fluorescence microscopy with a time binning of 20 ms (black line) and corresponding PL lifetime values derived from a running average of 500 ms stepped in 5 ms increments (red dots). Two-step photobleaching occurs. (b) A 4.6 Å dimer. Irreversible single-step bleaching occurs after ~6 s. This single (close spacing) or two-step (wide spacing) bleaching behaviour is representative of 70 % of the molecules studied from each group. Temporal fluctuations in $\tau_{PL}$ are much larger than in panel (a). The maximum difference in PL lifetime is denoted as $\Delta_{max}\tau_{PL}$ and is only defined during continuous PL emission, where both oligomers of the dimer are photoactive.

From the static snapshot of the $\tau_{PL}$ versus $E_{0-0}$ correlation in Fig. 1 it is not clear whether interchromophoric coupling varies from dimer to dimer, or fluctuates in each dimer with time. Since the relaxed exciton wavefunction is smaller than the conjugated segment of the dimer,[43] the exciton can localize to different regions of the molecule[27] which should impact the interchromophoric coupling efficiency. To resolve this question, we performed separate experiments in which the entire PL intensity from one molecule was directed onto one avalanche photodiode. This approach allowed us to record PL intensity traces with a sufficient number of photons to compute $\tau_{PL}$ decays within a time window of 500 ms. PL transients from 304 molecules of the 4.6 Å dimer and 125 molecules of the 21 Å dimer were selected under the condition that the PL intensity remained stable over at least 3 s before irreversible photobleaching occurred. Typical PL intensity traces, binned over 20 ms time intervals, are shown in Fig. 2 a) (21 Å dimer) and Fig. 2 b) (4.6 Å dimer). Whereas in this example the 21 Å dimer exhibits *two-step* bleaching after ~7 s and ~10.5 s, the 4.6 Å dimer bleaches in one step after ~6 s. This bleaching behaviour – single step for the narrow dimer, two step for the broad dimer – is typical, occurring in ~70 % of the 429 molecules studied, and provides a further differentiation between weak and strong interchromophoric coupling.[44] For weakly coupled chromophores, bleaching occurs independently, whereas for strongly-coupled units the chromophores behave as one.[28,32,45] The corresponding $\tau_{PL}$ values are superimposed as red dots in Fig. 2 and were extracted as a running average over the 500 ms integration window of $\tau_{PL}$ in steps of 5 ms. This running average allows us to extract the lowest and highest $\tau_{PL}$ values for each molecule. The maximum variation in $\tau_{PL}$ ($\Delta_{max}\tau_{PL}$) is derived during stable PL as denoted in Fig. 2. This approach ensures that $\tau_{PL}$ is only considered for periods during which both oligomers are photoactive. The PL transients show large fluctuations of $\tau_{PL}$ over several nanoseconds for the 4.6 Å dimer; virtually no fluctuations are seen for the 21 Å dimer.

By splitting the PL at an energy of ~2.66 eV into green and blue components and detecting the two beams with two avalanche photodiodes, we confirmed that a reduction in $\tau_{PL}$ coincides with a blue-shift of the PL spectrum (see Fig. S7). Interestingly, in this example, almost no change in the overall PL intensity is observed as $\tau_{PL}$ fluctuates. Typically, as seen in Fig. 1 e), formation of the excimer-like state coincides with a reduction in fluorescence yield. However, if the PL quantum yield of the individual oligomer is close to unity to start with the PL intensity should not fluctuate, since electronic dimerization only affects the radiative rate and the non-radiative rate is negligible to begin with.[40] The ensemble solution PL quantum yield was measured to be ~65 % for all materials (see ref.[28] for experimental details), but this value likely scatters between individual molecules.

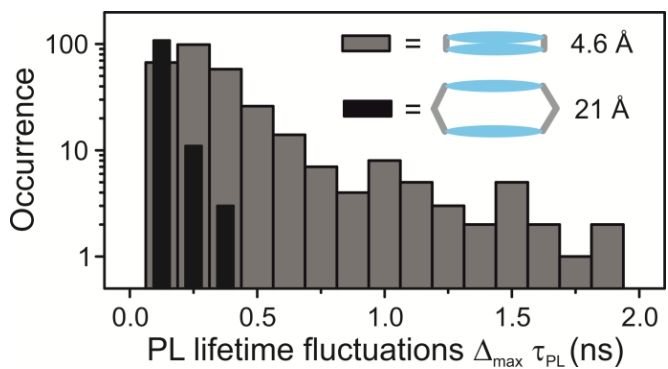

**Figure 3.** Histograms of the maximum PL lifetime fluctuations ($\Delta_{max}\tau_{PL}$) for 304 molecules of 4.6 Å spacing (grey bars) and for 125 molecules of 21 Å (black bars) spacing.

To test for the underlying dynamics in the formation of the excimer-like state, $\Delta_{max}\tau_{PL}$ values were extracted from a collection of transients. These values are plotted in a histogram in Fig. 3 for the 4.6 Å (grey bars) and 21 Å (black bars) dimers. A broad distribution with values ranging up to 1.8 ns is found for the narrow, but not for the wide dimers. We conclude that these fluctuations are solely related to dynamics in *interchromophoric* coupling and do not arise from dielectric changes in the environment, which would also influence the monomeric transition dipole moment.[46]

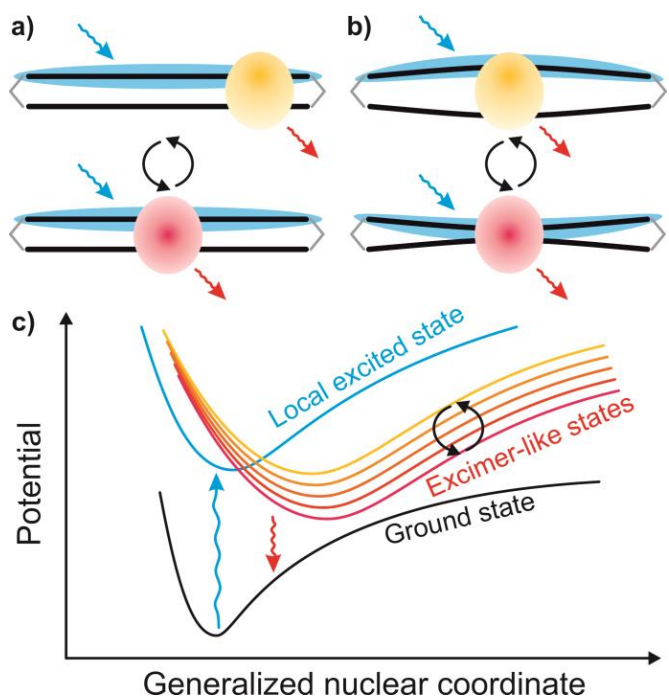

**Figure 4.** Qualitative model to describe the origin of fluctuations in excited-state interchromophoric (intramolecular) coupling strength due to either temporal variations in exciton localization or structural changes of the dimer. (a) One oligomer (black line) is excited by a photon (blue arrow) from the ground to a local excited state (blue). The exciton can localize at different parts of the oligomer, leading to different degrees of interchromophoric coupling in the excited state (red: stronger coupling; yellow: weaker coupling). (b) Molecular motion may also induce slight changes in interchromophoric spacing and coupling strength. The effect of those fluctuations in coupling strength is sketched in (c) with the ground state (black curve), the local (monomeric) excited state (blue curve) and different excimer-like states (yellow to red). Note

that for emission to occur, dimerization must be weak – a perfect excimer has a repulsive ground state making it non-emissive.

Dynamics can arise either due to spatial fluctuations of the position of exciton localization in the π-conjugated system,[27,43] or due to molecular motion including rotation of the phenylene units. Aggarwal *et al.* reported that the exciton can localize randomly on different parts of an extended π-conjugated system.[27] The basic idea is sketched in Fig. 4 a). The monomeric exciton is exposed to more π-electrons from the second cofacial oligomer if it localizes in the center (Fig. 4 a, top) as compared to the edges (Fig. 4 a, bottom). Localization in the center of one oligomer leads to stronger interchromophoric coupling and the emergence of an energetically lower-lying excimer-like state (red curve in Fig. 4 c) than localization at the edges (yellow curve in Fig. 4 c). We note that, strictly, it is not possible to separate mechanical molecular motion of the dimer segments and fluctuations in the region of exciton localization: exciton self-trapping is fundamentally linked to molecular dynamics.[43] This exciton localization process is also responsible for the main difficulties in calculating the coupling strength between elongated π-conjugated segments. One would need to perform TDDFT calculations in the excited state of these model systems, which is beyond the scope of this manuscript. The alternative explanation in Fig. 4 b) for the fluctuating interchromophoric coupling therefore invokes slight changes in interchromophoric spacing. Excitonic dimerization is more pronounced if the chromophores are closer together (Fig. 4 b, top). This spacing can vary with time, due to slight changes in the environment. Additionally, the phenylene units in the chromophores are relatively free to rotate at room-temperature in the 21 Å dimer, whereas this motion is inhibited in the 4.6 Å dimer. This might also impact the dynamics of interchromophoric coupling.

Well-defined model systems as shown here, in combination with single-molecule spectroscopy provide, the possibility to deconstruct a bulk molecular material, such as a conjugated polymer film, to the level of the very first intermolecular building block. Rather than a conventional intramolecular conjugated chromophore, in this case the spectroscopic unit resembles a physical dimer with excimer-like properties. We found that the average $\tau_{PL}$, which reports on interchromophoric coupling strength, increases from 0.6 ns to 1.1 ns from the largest to the smallest dimers. More importantly, the *scatter* of $\tau_{PL}$ values rises four-fold as molecular emission changes from monomeric to excimer-like, implying a strong spectral influence of molecular dynamics on the timescale of seconds. This fundamental dynamic heterogeneity in strong interchromophoric coupling, which amplifies non-ergodicity of single molecules, must be taken into account in formulating a microscopic understanding of the flow of excitation energy in bulk optoelectronic systems. For example, raising the excited-state lifetime without loss in quantum yield can be employed to increase exciton diffusion lengths in devices by inhibiting recombination.[20]

## ASSOCIATED CONTENT

**Supporting Information.** Details of experimental setup and further measurements. This material is available free of charge via the Internet at http://pubs.acs.org.

## AUTHOR INFORMATION

**Corresponding Author**

*Email: jan.vogelsang@physik.uni-regensburg.de

**Notes**

The authors declare no competing financial interests.

ACKNOWLEDGMENT

The authors are indebted to the European Research Council for funding through the Starting Grant MolMesON (305020) and to the Volkswagen Foundation for continued support of the collaboration.